\documentstyle[12pt]{article}
\newcommand{\EQ}{\begin{equation}}
\newcommand{\EN}{\end{equation}}
\newcommand{\bea}{\begin{eqnarray}}
\newcommand{\ena}{\end{eqnarray}}
\newcommand{\dw}[1]{\Delta{\cal W}_{#1}}
\newcommand{\om}[1]{{\cal H}_{#1}}
\newcommand{\w}[1]{{\cal W}_{#1}}

\def\sqr#1#2{{\vcenter{\vbox{\hrule height.#2pt
     \hbox{\vrule width.#2pt height#1pt \kern#1pt
           \vrule width.#2pt}
       \hrule height.#2pt}}}}

\begin{document}
\begin{flushright}
\begin{minipage}{0.25\textwidth} hep-th/0007222
\end{minipage}
\end{flushright}

\begin{center}
\bigskip\bigskip\bigskip
{\Large{\bf Consistency conditions and trace anomalies\\  in six dimensions}}

\vskip 2cm
F. Bastianelli, G. Cuoghi and  L. Nocetti 
\vskip .3cm
{\em  Dipartimento  di Fisica, Universit\`a di Bologna \\ and 
 INFN, Sezione di Bologna\\
via Irnerio 46, I-40126 Bologna, Italy} \\[.4cm]
\end{center}

\baselineskip=18pt

\vskip 3cm
\centerline{\large{\bf Abstract}}
\vskip .8cm

Conformally invariant quantum field theories develop trace anomalies 
when defined on curved backgrounds. 
We study again the problem of identifying all possible trace anomalies 
in $d=6$ by studying the consistency conditions to derive their 10 
independent solutions.
It is known that only 4 of these solutions represent true anomalies,
classified as one type A anomaly, given by the topological Euler density,
and three type B anomalies, made up by three independent Weyl invariants.
However, we also present the explicit expressions of the remaining 
6 trivial anomalies, namely those that can be obtained by the Weyl 
variation of local functionals.
The knowledge of the latter is in general necessary to disentangle the
universal coefficients of the type A and B anomalies from calculations 
performed on concrete models.

\newpage
\bigskip\bigskip\bigskip
\section{Introduction}

Six-dimensional conformal field theories (CFT$_6$) have attracted 
some interest in view of recent advances in string/M-theory.
In particular, the low energy dynamics of the collective coordinates of 
$N$ coinciding M5 branes of M-theory realize a quite interesting class
of non-trivial CFT$_6$ with maximal supersymmetry, the so-called 
${\cal N}=(0,2)$ interacting theories originally described 
in \cite{Wit,Stro}. 
While they still lack a lagrangian formulation, the AdS/CFT conjecture 
\cite{Mal,GKP,Wit2,rev} has provided concrete tools to extract 
some informations on these rather mysterious theories
in their large $N$ limit, such as their trace anomalies \cite{HS}, 
the spectrum of their operators \cite{AOY}
and some of their 2- and 3-point correlation functions \cite{AF,CFN,BZ}.
To understand better the structure of these interacting theories, 
in refs. \cite{BFT1,BFT2} some of their more accessible properties
were compared with those of another maximally supersymmetric
CFT$_6$: the non-interacting one made up by $N$ copies of the free 
${\cal N}=(0,2)$ tensor multiplet 
containing 5 scalars, 1 two-form with selfdual field strength
and 2 Weyl fermions. In particular, in \cite{BFT2}
the trace anomalies for the free theory were computed and compared with
the corresponding ones for the interacting theory obtained through
the AdS/CFT conjecture \cite{HS}. 
In this comparison it is crucial to disentangle 
the coefficients of the universal part of the anomalies by
separating out a trivial sector. The latter can always be cancelled by the 
variation of a local counterterm which can be
added to the effective action. 
This disentanglement
will be the subject of the present paper and we will start 
addressing it after a brief introduction to the topic of trace anomalies.

Trace anomalies can be characterized by the anomalous Weyl variation of a 
general coordinate invariant effective action depending on a 
background metric. Discovered originally in \cite{dis}
(and reviewed in \cite{venti}) they can be computed
using Feynman graphs, as in the original papers, or more efficiently
using the heat kernel methods of De Witt \cite{dewitt} 
(as employed e.g. in \cite{Bir,FrTs}) or by a quantum mechanical 
representation as proposed in \cite{FB}.
Their structure has been analyzed in \cite{Bon,Bon2,kar} 
by cohomological methods which encode the information on the 
Wess--Zumino consistency conditions \cite{WesZum} 
specialized to the Weyl symmetry.
Finally, a useful classification was described in \cite{DesSch}
where trace anomalies are divided into three classes:
type A (always proportional to the topological Euler density),  
type B (made up by independent Weyl invariants)
and  trivial anomalies (obtainable as Weyl variations of local functionals
and in general expressible as total derivatives).
The number of type B and trivial anomalies grows 
quite rapidly with the number of dimensions. Already in six dimensions
there are 1 type A, 3 type B and 6 trivial anomalies  satisfying
the consistency conditions on a set of 17 independent terms
with the property of being cubic in the curvature (and thus with the 
correct dimensions to constitute possible trace anomalies).
Usually a concrete calculation on a specific model
delivers the anomaly as a linear 
combination of these 17 terms, and one is left with the problem
of disentangling the correct universal part of the anomaly (type A and B).

This problem was solved pragmatically in \cite{BFT2} by expressing 
the $6d$ trace anomalies in a special basis for the curvature invariants 
(a basis which employs the Weyl tensor and traceless Ricci tensor instead 
of the Riemann and Ricci tensors).
This special basis makes it easier to cast the anomaly into a form with the 
expected type A and B contributions plus a combination of total derivatives.
The latter part was interpreted as a trivial anomaly since on general 
grounds one expects trivial anomalies to be total derivatives. 
This procedure worked for the four cases of free scalars, fermions, 
two-forms and interacting $(0,2)$ theory considered in \cite{BFT2}.
Now, one may object that the calculations made there used a basis 
of 7 independent total derivatives while the cohomological analysis 
of \cite{Bon2} predicts only 6 trivial anomalies. 
Thus, to make sure of the correct identification of the various anomalies,
we have decided to perform again a cohomological 
analysis to derive a basis for the trivial anomalies.
This allows us to check that indeed there are only 6 trivial anomalies,
for which we find the explicit expressions and identify
the local counterterms that can cancel them. Then we verify
that one specific linear combination of the 7 total derivative 
terms used in \cite{BFT2} doesn't solve the consistency conditions,
but never appears in the results for the trace anomalies 
of the various cases treated there, thus confirming 
the correctness of those results.

The final output of our analysis is a systematic classification
of the type A, type B and trivial anomalies for trace anomalies in 
six dimensions. This knowledge can be useful to put new calculations of 
such anomalies in a preferred basis and extract
unambiguously the universal coefficients of the type A and B parts.
It is presumably the difficulties related to the proper 
factorization of trivial anomalies that has caused a miscalculation 
of the trace anomalies for a scalar field in \cite{Ichi}.

Thus in sect. 2 we review and solve the consistency conditions.
In sect. 3 we cast those solutions into a more useful basis by taking 
into account their character as anomalies of type A, B and trivial.
In sect. 4 we present our conclusions. Finally, we leave 
appendices A, B, C and D for more technical parts where we
list useful results of our calculations.

\section{Consistency conditions}

Let's consider the effective action $W[g]$ for a CFT coupled to a 
background metric $g_{ab}$. We will use an euclidean signature.
For simplicity we assume absence of chiral gravitational anomalies
and thus can consider $W[g]$  to be general coordinate invariant. 
This assumption is not necessary \cite{Bon2}.
Under an infinitesimal Weyl transformation depending on an infinitesimal 
arbitrary function $\sigma(x)$
\bea 
\delta_\sigma g_{ab}(x) =2 \sigma(x) g_{ab}(x) ,
\ena
the effective action generically suffers an anomalous variation
\bea
\delta_\sigma W[g] =\int d^6x\ \sqrt{g} \sigma(x) A(x) .
\label{anomalia}
\ena
It is well known that by functional differentiation with respect to
${2\over \sqrt{g}}{\delta\over \delta g_{ab}}$
the effective action $W[g]$ generates correlation functions
of the stress tensor $T^{ab}$. 
Thus eq.~(\ref{anomalia}) produces an anomalous 
trace to the stress tensor
\bea
\langle T^a{}_a (x) \rangle= A(x)
\ena
which depends on the background curvature. 
General coordinate invariance guarantees that the anomaly $ A(x)$
is a scalar and dimensional considerations in $d=6$ fix it to be cubic 
in the curvature (two covariant derivatives count as one curvature).
However, those particular anomalies that can be obtained also from the Weyl
variation of local functionals of the metric are considered trivial since  
they can be cancelled by subtracting the same local functionals from 
the original non-local effective action.

Since the anomaly $ A(x)$ is obtained by varying a functional, there are 
integrability conditions that can be identified by applying the 
commutator algebra of the Weyl symmetry,
$[\delta_{\sigma_1},\delta_{\sigma_2}] =0$,
to the effective action.
Such integrability conditions are generically known as Wess--Zumino  
consistency conditions, originally derived for chiral anomalies 
in \cite{WesZum}.

Now, we follow the work of Bonora et al. \cite{Bon2} 
as a guideline to study the  consistency conditions 
\bea
[\delta_{\sigma_1},\delta_{\sigma_2}] W[g] =0
\label{eqWZ}
\ena
and derive all of their solutions in $d=6$.
We use the following conventions for the  curvature tensors
\bea
 [ \nabla_a, \nabla_b] V^c = R_{ab}{}^c{}_d V^d  , \ \ \ 
R_{ab}= R_{ca}{}^c{}_b  ,\ \ \ \  R= R^a{}_a ,
\ena
so that the scalar curvature of a sphere is positive, 
and use the same basis of 17 independent curvature invariants
as in \cite{Bon2}\footnote{
However we differ in the definitions of the various curvature
tensors, so there are some sign differences with respect to 
ref.~\cite{Bon2}. In particular, we agree on the sign of the Riemann
tensor $R_{abcd}$ but 
have opposite sign for the Ricci tensor $R_{ab}$
and scalar curvature $R$.}  
\EQ
\begin{array}{lll}
K_1 = R^3  & 
K_2 = R R_{ab}^2 & 
K_3 = R R_{abmn}^2 \\ [3mm] 
K_4 = R_a{}^m R_m{}^i R_i{}^a & 
K_5 =  R_{ab} R_{mn} R^{mabn} & 
K_6= R_{ab} R^{amnl} R^b{}_{mnl} \\ [3mm] 
K_7 = R_{ab}{}^{mn} R_{mn}{}^{ij}R_{ij}{}^{ab} \ \ \ \ &
K_8 = R_{amnb} R^{mijn} R_{i}{}^{ab}{}_j \ \ \ \ &
K_9 = R\nabla^2 R \\ [3mm]
K_{10} = R_{ab}\nabla^2 R^{ab} &
K_{11} = R_{abmn}\nabla^2 R^{abmn} &
K_{12} = R^{ab} \nabla_a \nabla_b R \\ [3mm]
K_{13} = (\nabla_a R_{mn})^2  &
K_{14} = \nabla_a R_{bm} \nabla^b R^{am} &
K_{15} = (\nabla_i R_{abmn})^2  \\ [3mm]
K_{16} = \nabla^2 R^2  &
K_{17} =\nabla^4 R . & 
\label{inv}
\end{array}
\EN 
All other terms cubic in the curvature are linear combinations of 
the above invariants after 
taking into account the symmetry 
properties and the Bianchi identities of the Riemann tensor. 

Any trace anomaly can be expanded in the above basis
\bea
\delta_\sigma W[g] =\int d^6x\ \sqrt{g} \sigma(x) \sum_{i=1}^{17} a^i K_i 
\label{sei}
\ena
and after computing a second Weyl variation one obtains
\bea
[\delta_{\sigma_2},\delta_{\sigma_1}] W[g] 
=\int d^6x\ \sqrt{g} \sum_{i=1}^{17} 
\sum_{\alpha=1}^{9}  f^\alpha{}_i a^i H_\alpha 
\label{WZcc}
\ena 
where the rectangular matrix of coefficients $f^\alpha{}_i$
can be constructed using the variations of the terms
entering eq.~(\ref{sei}) and reported in appendix A, and where
the 9 independent (unintegrated) 2-cochains $H_\alpha$ are given 
by\footnote{The symbol
$[\cdot \cdot ]$ denotes antisymmetrization, 
namely $a_{[1} b_{2]} = a_1 b_2 - a_2 b_1$.}
\EQ 
\begin{array}{ll}
H_1= R^2 \sigma_{[1} \nabla^2 \sigma_{2]} &
H_2= R_{ab}^2  \sigma_{[1} \nabla^2 \sigma_{2]}   \\ [3mm]
H_3= R R_{ab} \sigma_{[1} \nabla^a \nabla^b \sigma_{2]} & 
H_4= R_{am} R^m{}_b \sigma_{[1} \nabla^a \nabla^b \sigma_{2]} \\ [3mm] 
H_5=  R_{abmn}^2 \sigma_{[1} \nabla^2 \sigma_{2]} &
H_6= (\nabla^2 R) \sigma_{[1} \nabla^2 \sigma_{2]}   \\ [3mm]
H_7= R \sigma_{[1} \nabla^4 \sigma_{2]} &
H_8= R_{ab} \sigma_{[1} \nabla^a \nabla^b \nabla^2 \sigma_{2]}  \\ [3mm]
H_9= R^{mn} R_{amnb} \sigma_{[1} \nabla^a \nabla^b \sigma_{2]} .
 \ \ \ \ & \\ [3mm]  
\end{array}
\label{2c}
\EN

Now the consistency condition eq.~(\ref{eqWZ}) 
applied to eq.~(\ref{WZcc}) requires that
\bea
\sum_{i=1}^{17}  f^\alpha{}_i\  a^i =0 , \hskip 1cm \alpha=1,\dots,9.
\label{10}
\ena
The $9 \times 17$ matrix $ f^\alpha{}_i $ has rank 7, so there are
7 independent constraints for the coefficients $a^i$
to form a consistent anomaly. The resulting    
10 independent anomalies can be presented as
\bea
M_I(x) = \sum_{i=1}^{17} a_I^i K_i(x) 
\ena
where the 10 vectors $a_I^i$,  $I=1,\dots,10$, 
form a basis for the solutions of eq. (\ref{10}). 
These vectors are constructed in appendices A 
(where one can read off the matrix $f^\alpha{}_i$) and C.
We list them later on in eqs. (\ref{a}--\ref{b}).

Trivial anomalies are those that can be obtained by varying
a local functional. To recognize them we compute the Weyl variation
of the most general local functional obtained as a linear combination
with coefficients $c^i$ of the integrated curvature invariants $K_i$ 
(note that we can restrict the index $i\leq 10$ since by partial 
integration the remaining terms are not linearly independent)
\bea
\delta_\sigma 
\int d^6x\ \sqrt{g} 
\sum_{i=1}^{10} c^i K_i 
= \int d^6x\ \sqrt{g} \sigma(x)
\sum_{i=1}^{10} \sum_{j=1}^{17}
g^j{}_i c^i K_j .
\label{triv}
\ena
The matrix of coefficients $g^j{}_i$ has rank 6 and therefore identifies  
6 trivial anomalies. These are constructed in appendices B 
(where one can read off the matrix $g^j{}_i$) and C, and 
reported here below in eqs. (\ref{c}--\ref{b}).
In appendix C one may also find the local functionals  which 
generate the trivial anomalies (see eq. (\ref{locfun})).

Now we present the solutions of the consistency conditions
just described.
A basis for the non-trivial anomalies is given by
\bea
M_{1} &=& \frac{19}{800}K_{1}-\frac{57}{160}K_{2}+\frac{3}{40}K_{3}+
  \frac{7}{16}K_{4} -\frac{9}{8}K_{5}-\frac{3}{4}K_{6}+K_{8} 
\label{a}
\\ [2mm]
M_{2} &=& \frac{9}{200}K_{1}-\frac{27}{40}K_{2}+\frac{3}{10}K_{3}+
  \frac{5}{4}K_{4}-\frac{3}{2}K_{5}-3K_{6}+K_{7} \\ [2mm]
M_{3} &=& -K_{1}+8K_{2}+2K_{3}-10K_{4}+10K_{5}-\frac{1}{2}K_{9}+5K_{10}-
      5K_{11} \\ [2mm]
M_{4} &=& -K_{1}+12K_{2}-3K_{3}-16K_{4}+24K_{5}+24K_{6}-4K_{7}-8K_{8}
\ena
where we have chosen to agree with the ones reported in ref.~\cite{Bon2}.
Instead a suitable basis for the remaining trivial anomalies is given by 
\bea
M_{5} &=& 6K_{6}-3K_{7}+12K_{8}+K_{10}-7K_{11}-11K_{13}+12K_{14} -4K_{15} 
\label{c}
\\ [2mm]
M_{6} &=& -\frac{1}{5}K_{9}+K_{10}+\frac{2}{5}K_{12}+K_{13} \\ [2mm]
M_{7} &=& K_{4}+K_{5}-\frac{3}{20}K_{9}+\frac{4}{5}K_{12}+K_{14} 
\\ [2mm]
M_{8} &=& -\frac{1}{5}K_{9}+K_{11}+\frac{2}{5}K_{12}+K_{15} \\ [2mm]
M_{9} &=& K_{16} \\ [2mm]
M_{10} &=& K_{17}.
\label{b}
\ena
This is the main result we were searching for.

\section{A useful basis for six dimensional trace anomalies}

In the previous section we have derived the solutions to the
consistency conditions. 
We now put those solutions into a more useful basis by taking into 
account their character as type A, B or trivial anomalies, as 
classified in \cite{DesSch}.

The type A anomaly is unique and proportional to the   
six dimensional topological Euler density and can be written as
\bea
E_{6} &=& -\epsilon_{m_1n_1m_2n_2m_3n_3}\epsilon^{a_1b_1a_2b_2a_3b_3}
          R^{m_1n_1}{}_{a_1b_1}R^{m_2n_2}{}_{a_2b_2} R^{m_3n_3}{}_{a_3b_3}
          =8M_{4} .
\ena
The anomalies of type B are given instead by the three following Weyl 
invariants
\bea
I_{1} &=& C_{amnb} C^{mijn} C_{i}{}^{ab}{}_j=M_{1} \\
I_{2} &=& C_{ab}{}^{mn} C_{mn}{}^{ij} C_{ij}{}^{ab}=M_{2} \\
I_{3} &=& C_{mabc}\left(\nabla^{2}\delta^{m}_n+4R^{m}_n
-\frac{6}{5}R\delta^{m}_n\right)C^{nabc}+\nabla_{i}{ J}^{i} \cr
      &=& \frac{16}{3}M_{1}+\frac{8}{3}M_{2}-\frac{1}{5}M_{3}+
\frac{2}{3}M_{4}+\nabla_{i}{ J}^{i}
\ena
where
\EQ
C_{abcd}=R_{abcd}-\frac{1}{4}(g_{ac}R_{bd}+g_{bd}R_{ac}-g_{ad}R_{bc}
-g_{bc}R_{ad})+\frac{1}{20}(g_{ac}g_{bd}-g_{ad}g_{bc})R
\label{Weyltensor}
\EN
is the Weyl tensor in 6 dimensions and
\bea
\nabla_{i}{ J}^{i}=-\frac{2}{3}M_{5}-\frac{13}{3}M_{6}+2M_{7}
+\frac{1}{3}M_{8}
\ena
is a trivial anomaly that make $I_{3}$ locally Weyl invariant
once multiplied by the measure $\sqrt g$ \cite{Erdmenger:1997gy}.
Finally the independent six trivial anomalies can be identified by 
$M_5$, $M_6$, $M_7$, $M_8$, $M_9$, $M_{10}$ as 
listed in eqs.~(\ref{c}--\ref{b}).

To summarize, a preferred basis for the trace anomalies
which takes into account the classification of ref.~\cite{DesSch}
is given by $(E_6; I_1, I_2, I_3; M_5, M_6, M_7, M_8, M_9, M_{10})$.
The first four elements make up a basis for the true trace anomalies
and in that form have been used in the calculations of ref. \cite{BFT2}.

It may be useful to recall that in the basis of Bonora et al.~\cite{Bon2} 
$M_4$ gives the type A anomaly, while $M_1$ and $M_2$ 
are type B anomalies. On the other hand, 
$M_3$ contains a spurious contribution
from the Euler density and it is not classifiable as the 
remaining type B anomaly: it is preferable to use $I_3$ instead.

Anselmi has introduced in ref.~\cite{Anselmi:2000uk} the notion 
of pondered Euler density ${\tilde E_6}$ by adding a suitable 
trivial anomaly to ${ E_6}$  to make it linear in 
the conformal factor once evaluated on conformally flat metrics
\bea
\tilde E_6 = E_6 +\biggl ({288\over 5} - 20 \zeta \biggr )M_6
+\biggl( 20\zeta -{408\over 5} \biggr) M_7 +
\biggl ({\zeta\over 2} -{9\over 25} \biggr )M_9 -{24\over 5} M_{10} .
\ena
This is an equivalent way of presenting the type A anomaly which
may be useful for various applications. 
Note that $\zeta$ labels a 1-parameter family of trivial anomalies.
It can be chosen at will showing that the definition of a 
pondered density is not unique.
This construction can be extended to any even 
dimension \cite{Anselmi:2000uk}. 

A further characterization 
of type A anomalies has been proposed in \cite{imb}
by studying the AdS/CFT holographic correspondence, while
their systematic computation for free models
in arbitrary dimensions has been carried out recently in \cite{cap}.
Finally, it is worth mentioning that CFTs with a special linear relation
between the type A and B anomalies, the so-called $c=a$ theories,
have been identified in \cite{ans3} as an interesting subclass of 
conformal theories with special properties. 

\section{Conclusions}

We have presented a systematic derivation and classification
of trace anomalies in six dimensions.
We have solved the consistency conditions and listed the 10 independent
solutions as type A ($E_6$), type B 
($I_1,\ I_2,\ I_3$) and trivial ($M_5,\ M_6,\ M_7,\ M_8,\ M_9,\ M_{10}$)
trace anomalies. We summarize them by using the $K_i$ basis
in Table 1.

Our motivation to perform this analysis was to make sure 
that the identifications of type  A and B  anomalies 
made in \cite{BFT2} for various models was correct.
As we show explicitly in appendix D that is the case:
a spurious term, which doesn't solve the consistency conditions
but enters the basis of total derivatives used 
to identify trivial anomalies, always drops out in the relevant cases.
On the other hand, the calculation of the trace anomalies 
for a scalar field performed in 
\cite{Ichi} didn't produce the same result as in \cite{BFT2}
because trivial anomalies were not properly factorized:
in \cite{Ichi} the coefficients of $K_4$, $K_5$, $K_7$ and $K_8$ 
were interpreted
as the coefficients of true trace anomalies but those structures 
appear in our list of trivial anomalies,
so their coefficients do not have a universal meaning
since they can be modified by adding local counterterms to the 
effective action. Nevertheless, the method of \cite{Ichi} can 
still be used to compute trace anomalies. By inspecting Table 1  
one can recognize the terms
that are not corrupted by trivial anomalies: a simple set of
4 independent structures is given by $K_1$, $K_2$, $K_3$, $K_4 -K_5$.
 
With the present knowledge of trivial anomalies at hand, an interesting
investigation could be to study their flows in CFT$_6$  deformed by  
relevant operators. This would provide a test in six dimensions 
of some of the properties studied in ref. \cite{Anselmi:1999xk}.

\begin{table}
\begin{center}
\[\begin{array}{|l|} \hline \\ [-3.2mm]
E_6 =   -8 K_1 + 96 K_2 - 24K_3 - 128K_4 + 192 K_5 +192 K_6 
-32 K_7 -64 K_8 \\ [1mm]  \hline\hline \\[-3.2mm]
I_1 = \frac{19}{800}K_{1}-\frac{57}{160}K_{2}+\frac{3}{40}K_{3}+
  \frac{7}{16}K_{4} -\frac{9}{8}K_{5}-\frac{3}{4}K_{6}+K_{8} 
\\[1mm]  \hline \\ [-3.2mm]
I_2 = \frac{9}{200}K_{1}-\frac{27}{40}K_{2}+\frac{3}{10}K_{3}+
  \frac{5}{4}K_{4}-\frac{3}{2}K_{5}-3K_{6}+K_{7} 
\\ [1mm] \hline  \\ [-3.2mm]
I_3 = - {11\over 50}K_{1}+{27\over10}K_{2}-{6\over5}K_{3}
-K_{4}+ 6 K_{5} +2 K_7 -8K_8
+\frac{3}{5}K_{9}-6K_{10}+6K_{11}+3K_{13}-6K_{14} +3 K_{15} 
\\ [1mm] \hline\hline \\ [-3.2mm]
M_5 =  6K_{6}-3K_{7}+12K_{8}+K_{10}-7K_{11}-11K_{13}+12K_{14} -4K_{15} 
\\ [1mm] \hline \\ [-3.2mm]
M_{6} = -\frac{1}{5}K_{9}+K_{10}+\frac{2}{5}K_{12}+K_{13} 
\\ [1mm] \hline \\ [-3.2mm]
M_{7} = K_{4}+K_{5}-\frac{3}{20}K_{9}+\frac{4}{5}K_{12}+K_{14} 
\\ [1mm] \hline \\ [-3.2mm]
M_{8} = -\frac{1}{5}K_{9}+K_{11}+\frac{2}{5}K_{12}+K_{15} 
\\ [1mm] \hline\\ [-3.2mm]
M_{9} = K_{16} 
\\ [1mm] \hline\\ [-3.2mm]
M_{10} = K_{17}
\\ [1mm] \hline
\end{array}\]
\end{center}
\vspace{-0.5cm}
\caption{Trace anomalies in six dimensions: 
type A, type B and trivial anomalies \label{tab1}}
\end{table}

\vfill\eject 

\section*{Appendix A}

We define the 1-cochains
\bea
{\cal W}^{(a)}_{i}= \int d^6x\ \sqrt{g} \sigma_a(x)K_{i}, 
\ \ \ \ \ \ \ i=1,\ldots, 17
\ena
where $\sigma_a$ denote infinitesimal parameters of 
Weyl transformations and $K_i$ belong to the list 
of curvature invariants given in eq.~(\ref{inv}),
and the  2-cochains
\bea
{\cal H}_{\alpha}= \int d^6x\ \sqrt{g} H_{\alpha}, \ \ \ \ \ \ \ 
\alpha =1,\ldots, 9
\ena
with the list of $H_{\alpha}$ reported in eq.~(\ref{2c}).
Now, we compute the variations
\bea
\dw{i} \equiv 
\delta_{\sigma_2} {\cal W}^{(1)}_{i}-\delta_{\sigma_1}{\cal W}^{(2)}_{i}
\ena
which can be expanded in the basis of the functionals ${\cal H}_{\alpha}$ .
We find:
\EQ
\begin{array}{lll}
\dw{1}&=&-30\om{1} \\ [2mm]
\dw{2}&=&-2\om{1}-10\om{2}-8\om{3} \\ [2mm]
\dw{3}&=&-8\om{3}-10\om{5} \\ [2mm]
\dw{4}&=&-3\om{2}-12\om{4} \\ [2mm]
\dw{5}&=&2\om{2}+2\om{3}-2\om{4}-8\om{9} \\ [2mm]
\dw{6}&=&-\om{1}+4\om{2}+4\om{3}-12\om{4}-2\om{5}+12\om{9} \\ [2mm]
\dw{7}&=&-3\om{1}+12\om{2}+12\om{3}-24\om{4}-3\om{5}+24\om{9} \\ [2mm]
\dw{8}&=&-\frac{3}{4}\om{1}+3\om{2}+3\om{3}-6\om{4}-\frac{3}{4}\om{5} \\ [2mm]
\dw{9}&=&-2\om{1}-10\om{6}-10\om{7} \\ [2mm]
\dw{10}&=&\om{1}-6\om{2}-4\om{3}-4\om{4}-3\om{6}-\om{7}
-4\om{8}-16\om{9} \\ [2mm]
\dw{11}&=&4\om{1}-12\om{2}-16\om{3}+16\om{4}-4\om{5}
-2\om{6}-4\om{8}-32\om{9} \\ [2mm]
\dw{12}&=&-\om{1}-5\om{6}-10\om{8} \\ [2mm] 
\dw{13}&=&-\om{1}+6\om{2}+4\om{3}+4\om{4}+3\om{6}-\om{7}+8\om{8}+16\om{9} \\ [2mm]
\dw{14}&=&\frac{1}{2}\om{1}+\om{2}-2\om{3}+14\om{4}
+\frac{5}{2}\om{6}-\frac{3}{2}\om{7}+8\om{8}+8\om{9} \\ [2mm]
\dw{15}&=&-4\om{1}+12\om{2}+16\om{3}-16\om{4}+4\om{5}
+2\om{6}-2\om{7}+8\om{8}+32\om{9} \\ [2mm]
\dw{16}&=& 0 \\ [2mm]                    
\dw{17}&=& 0 .
\end{array}
\EN
From this list one can read off the matrix $f^\alpha{}_i$ of 
eq.~(\ref{10}).

\section*{Appendix B}

To find all trivial anomalies, we compute the Weyl variation 
of the most general dimensionless local functional of the metric
\EQ
{\cal K}=\sum_{i=1}^{10}c^{i}{\cal K}_{i}\ \ \ \mbox{where}
\ \ \ {\cal K}_{i}=\int d^6x\ \sqrt{g}K_{i} .
\EN
Defining as before ${\cal W}_{i}= \int d^6x\ \sqrt{g} \sigma(x)K_{i}$ 
we have
\EQ
\begin{array}{lll}
\delta_{\sigma}{\cal K}_{1} &=& -30\w{16} \\ [2mm]
\delta_{\sigma}{\cal K}_{2} &=& 4\w{9}-20\w{10}-8\w{12}-20\w{13}
-6\w{16} \\ [2mm]
\delta_{\sigma}{\cal K}_{3} &=& 4\w{9}-20\w{11}-8\w{12}-20\w{15}
-4\w{16} \\ [2mm]
\delta_{\sigma}{\cal K}_{4} &=& -12\w{4}-12\w{5}+3\w{9}-6\w{10}
-12\w{12}-6\w{13}-12\w{14} -\frac{3}{2}\w{16} \\ [2mm]
\delta_{\sigma}{\cal K}_{5} &=& -10\w{4}-10\w{5}-4\w{6}+2\w{7}
-8\w{8}-\frac{1}{2}\w{9}+12\w{10}+2\w{11}-4\w{12} \\ [2mm] 
                       & & +20\w{13}-18\w{14}+\frac{3}{4}\w{16} \\ [2mm]
\delta_{\sigma}{\cal K}_{6} &=& 6\w{6}-3\w{7}+12\w{8}+\w{9}-4\w{10}
-7\w{11}-2\w{12}-16\w{13}+12\w{14}\\ [2mm]
                       & & -4\w{15}-\frac{1}{2}\w{16} \\ [2mm]
\delta_{\sigma}{\cal K}_{7} &=& 12\w{6}-6\w{7}+24\w{8}-12\w{11}
-24\w{13}+24\w{14}-6\w{15} \\ [2mm]
\delta_{\sigma}{\cal K}_{8} &=& -6\w{4}-6\w{5}+6\w{10}-\frac{3}{2}\w{11}
-3\w{12}+6\w{13}-6\w{14}-\frac{3}{2}\w{15} \\ [2mm]
\delta_{\sigma}{\cal K}_{9} &=& -2\w{16}-20\w{17} \\ [2mm]
\delta_{\sigma}{\cal K}_{10} &=& -20\w{4}-20\w{5}-8\w{6}+4\w{7}-16\w{8}
+3\w{9}+4\w{10}+4\w{11}-16\w{12}\\ [2mm]
                        & & +20\w{13}-36\w{14}-\frac{3}{2}\w{16}-6\w{17} .
\end{array}
\EN
Other variations are not necessary since by partial integration we find
\EQ
\begin{array}{lll}
{\cal K}_{11} &=& -4{\cal K}_{4}-4{\cal K}_{5}+2{\cal K}_{6}-{\cal K}_{7}+4{\cal K}_{8}-{\cal K}_{9}+4{\cal K}_{10} \\ [1mm]
{\cal K}_{12} &=& \frac{1}{2}{\cal K}_{9} \\ [1mm]
{\cal K}_{13} &=& -{\cal K}_{10} \\ [1mm]
{\cal K}_{14} &=& -{\cal K}_{4}-{\cal K}_{5}-\frac{1}{4}{\cal K}_{9} \\ [1mm]
{\cal K}_{15} &=& -{\cal K}_{11} \\ [1mm]
{\cal K}_{16} &=& {\cal K}_{17}=0 .
\end{array}
\EN
From the set of Weyl variations written above one can easily constructs 
the matrix $g^{j}{}_{i}$ of eq.~(\ref{triv}). This matrix has rank 6 
and so there are 6 independent trivial anomalies, which can be 
identified as the variations of 
${\cal K}_{1}$, ${\cal K}_{2}$, 
${\cal K}_{3}$, ${\cal K}_{5}$, ${\cal K}_{6}$, ${\cal K}_{9}$.

\section*{Appendix C}

Here we solve the consistency conditions in eq.~(\ref{eqWZ}).
They consist in the following homogeneous system of linear equations 
(same as eq.~(\ref{10}) of the main text)
\bea
\sum_{i=1}^{17}  f^\alpha{}_i\  a^i =0 , \hskip 1cm \alpha=1,\dots,9.
\ena
 The $9\times 17$ matrix $f^\alpha{}_{i}$ 
can be constructed form the calculations reported in appendix A.
It has rank 7 and so one has 10 independent solutions for the $a^{i}$. 
A possible choice for the parameters of the solution is: 
$a^{6}$, $a^{7}$, $a^{8}$, $a^{10}$, 
$a^{11}$, $a^{13}$, $a^{14}$, $a^{15}$, $a^{16}$, $a^{17}$. 
Thus, one can straightforwardly derive a suitable basis for the anomalies:
\EQ
\begin{array}{lll}
A_{1} &=& -\frac{21}{200}K_{1}+\frac{43}{40}K_{2}-\frac{1}{5}K_{3}
-\frac{5}{4}K_{4}+\frac{3}{2}K_{5}+K_{6} \\[2mm]
A_{2} &=& -\frac{27}{100}K_{1}+\frac{51}{20}K_{2}-\frac{3}{10}K_{3}
-\frac{5}{2}K_{4}+3K_{5}+K_{7}\\[2mm]
A_{3} &=& -\frac{11}{200}K_{1}+\frac{9}{20}K_{2}-\frac{3}{40}K_{3}
-\frac{1}{2}K_{4}+K_{8}\\[2mm]
A_{4} &=& \frac{3}{25}K_{1}-K_{2}-2K_{5}-\frac{1}{10}K_{9}+K_{10}
-\frac{2}{5}K_{12}\\[2mm]
A_{5} &=& \frac{8}{25}K_{1}-\frac{13}{5}K_{2}-\frac{2}{5}K_{3}+2K_{4}
-4K_{5}+K_{11}-\frac{2}{5}K_{12}\\[2mm]
A_{6} &=& -\frac{3}{25}K_{1}+K_{2}+2K_{5}-\frac{1}{10}K_{9}
+\frac{4}{5}K_{12}+K_{13}\\[2mm]
A_{7} &=& K_{4}+K_{5}-\frac{3}{20}K_{9}+\frac{4}{5}K_{12}+K_{14}\\[2mm]
A_{8} &=& -\frac{8}{25}K_{1}+\frac{13}{5}K_{2}+\frac{2}{5}K_{3}-2K_{4}
+4K_{5}-\frac{1}{5}K_{9}+\frac{4}{5}K_{12}+K_{15}\\[2mm]
A_{9} &=& K_{16}\\[2mm]
A_{10} &=& K_{17} .
\end{array} 
\EN
Now, in appendix B we have identified a basis of trivial anomalies.
One can use that knowledge to make a change of basis and separate
the 6 trivial anomalies from the non-trivial ones. We have chosen
the latter to agree with those of ref. \cite{Bon2} obtaining
\EQ
\begin{array}{l}
M_{1}=-\frac{3}{4}A_{1} +A_{3},\ \ \ \ 
M_{2}=-3A_{1}+A_{2},\ \ \ \ 
M_{3}=5A_{4}-5A_{5},\ \ \ \\[1mm]
M_{4}=24A_{1}-4A_{2}-8A_{3},\\[1mm] 
M_{5}=6A_{1}-3A_{2}+12A_{3}+A_{4}-7A_{5}-11A_{6}+12A_{7}-4A_{8},
\\[1mm] 
M_{6}=A_{4}+A_{6},\ \ \ \
M_{7}=A_{7},\ \ \ \
M_{8}=A_{8}+A_{5},\ \ \ \
M_{9}=A_{9},\ \ \ \
M_{10}=A_{10}.
\end{array}
\EN
This is the basis reported in eqs. (\ref{a}--\ref{b}).
The last six elements $M_5, \ldots, M_{10}$
are trivial and one can easily identify the local functionals 
generating them. Defining 
\EQ
{\cal M}_{i}= \int d^6x\ \sqrt{g} \sigma(x) M_{i},
\EN
we have
\EQ
\begin{array}{l}
 {\cal M}_5 = \delta_{\sigma} \left(\frac{1}{30}{\cal K}_1
-\frac{1}{4}{\cal K}_2+{\cal K}_6\right), \ \ 
 {\cal M}_6 = \delta_{\sigma}\left(\frac{1}{100}{\cal K}_1
-\frac{1}{20}{\cal K}_2\right), \ \ \\ [3mm]
 {\cal M}_7 = \delta_{\sigma}\left(\frac{37}{6000}{\cal K}_1
-\frac{7}{150}{\cal K}_2
+\frac{1}{75}{\cal K}_3-\frac{1}{10}{\cal K}_5
-\frac{1}{15}{\cal K}_6\right), \ \
 {\cal M}_8 = \delta_{\sigma}\left(\frac{1}{150}{\cal K}_1
-\frac{1}{20}{\cal K}_3\right), \ \ \\ [3mm]
 {\cal M}_9 = \delta_{\sigma}\left(-\frac{1}{30}{\cal K}_1\right), \ \
 {\cal M}_{10} = \delta_{\sigma}\left(\frac{1}{300}{\cal K}_1
-\frac{1}{20}{\cal K}_9\right). \ \
\end{array}
\label{locfun}
\EN
\section*{Appendix D}

In \cite{BFT2} (and  in \cite{PR}) 
the following basis of invariants was also used 
\EQ
\begin{array}{lll}
B_1 = \nabla^4 R & 
B_2 = (\nabla_{a} R)^2 & 
B_3 = (\nabla_a B_{mn})^2 \\ [2mm] 
B_4 = \nabla_a B_{bm} \nabla^b B^{am} & 
B_5 = (\nabla_i C_{abmn})^2 & 
B_6 = R\nabla^2 R \\ [2mm] 
B_7 = B_{ab}\nabla^2 B^{ab} &
B_8 = B_{ab} \nabla_m \nabla^b B^{am} &
B_9 = C_{abmn}\nabla^2 C^{abmn} \\ [2mm]
B_{10} = R^3  & 
B_{11} = R B_{ab}^2 & 
B_{12} = R C_{abmn}^2 \\ [2mm] 
B_{13} = B_a{}^m B_m{}^i B_i{}^a & 
B_{14} = B_{ab} B_{mn} C^{ambn} & 
B_{15} = B_{ab} C^{amnl} C^b{}_{mnl} \\ [2mm] 
B_{16} = C_{ab}{}^{mn} C_{mn}{}^{ij}C_{ij}{}^{ab}  &
B_{17} = C_{ambn} C^{aibj} C^{m}{}_{i}{}^{n}{}_{j} & \\ [2mm]
\end{array}
\EN 
where $C_{abcd}$ is the Weyl tensor defined in (\ref{Weyltensor}) 
and $B_{ab}$ is the traceless part of the Ricci tensor
\EQ
B_{ab}= R_{ab}-\frac{1}{6}R\, g_{ab}.
\EN
The trivial anomalies in \cite{BFT2} are expressed using the following
set of total derivatives \cite{PR}
\EQ
\begin{array}{l}
C_{1}=B_{1}, \ \ \ \ C_{2}=B_{2}+B_{6}, \ \ \ \ C_{3}=B_{3}+B_{7},
\ \ \ \ C_{4}=B_{4}+B_{8}, \ \ \ \ C_{5}=B_{5}+B_{9}, \\ [2mm]
C_{6}=\frac{1}{9}B_{2}-B_{4}-\frac{1}{5}B_{11}-\frac{3}{2}B_{13}+
B_{14}, \\ [2mm]
C_{7}=\frac{1}{60}B_{2}-\frac{3}{4}B_{3}+\frac{3}{4}B_{4}+\frac{1}{4}B_{5}
+\frac{1}{12}B_{12}+\frac{1}{2}B_{15}-\frac{1}{4}B_{16}-B_{17}.
\end{array}
\EN
However, these total derivatives do not form a subset of the space of 
trivial anomalies. In fact we find
\EQ
\begin{array}{l}
\hskip -.4cm 
C_{1}=M_{10}, \ \ \ C_{2}=\frac{1}{2}M_{9}, \ \ \ 
C_{3}=-\frac{1}{12}M_{9}+(K_{10}+K_{13}),\\[2mm] 
\hskip -.4cm 
C_{4}=-\frac{7}{6}M_{6}+M_{7}-\frac{5}{72}M_{9}+\frac{7}{6}(K_{10}+K_{13}), 
 \ \ \ C_{5}=-M_{6}+M_{8}+\frac{1}{20}M_{9}, \\ [2mm] 
\hskip -.4cm 
C_{6}=2M_{6}-M_{7}+\frac{1}{8}M_{9}-2(K_{10}+K_{13}),
\ \  C_{7}=\frac{1}{12}M_{5}-\frac{1}{12}M_{6}-\frac{1}{4}M_{7}
+\frac{7}{12}M_{8}+\frac{1}{32}M_{9}.
\end{array}
\EN
Because of the term $(K_{10}+K_{13})$ the space spanned by the $C_{i}$
is larger than the space of trivial anomalies. The latter is expressed by
\bea
 \sum_{i=1}^{7}a^{i}C_{i}\ \ \ \ \mbox{is a trivial anomaly iff}\ \ \ \ 
 6a_{3}+7a_{4}-12a_{6}=0 .
\ena
In \cite{BFT2} all the linear combinations of $C_{i}$ satisfy the above 
relation, i.e. are trivial anomalies.
This is a good check on the correctness of the anomaly computations
performed there.
\newpage



\begin{thebibliography}{99}
\baselineskip=16pt
\bibitem{Wit} 
E.~Witten,
``Some comments on string dynamics'',
{\tt hep-th/9507121};
``Five-branes and M-theory on an orbifold'',
Nucl.\ Phys.\  {\bf B463} (1996) 383,
{\tt hep-th/9512219}.

\bibitem{Stro} 
A.~Strominger,
``Open p-branes'',
Phys.\ Lett.\  {\bf B383} (1996) 44,
{\tt hep-th/9512059}.

\bibitem{Mal} 
J.~Maldacena,
``The large $N$ limit of superconformal field theories and supergravity'',
Adv.\ Theor.\ Math.\ Phys.\  {\bf 2} (1998) 231,
{\tt hep-th/9711200}.

\bibitem{GKP} 
S.~S.~Gubser, I.~R.~Klebanov and A.~M.~Polyakov,
``Gauge theory correlators from non-critical string theory'',
Phys.\ Lett.\  {\bf B428} (1998) 105,
{\tt hep-th/9802109}.

\bibitem{Wit2} 
E.~Witten,
``Anti-de Sitter space and holography'',
Adv.\ Theor.\ Math.\ Phys.\  {\bf 2} (1998) 253,
{\tt hep-th/9802150}.

\bibitem{rev} O.~Aharony, S.~S.~Gubser, J.~Maldacena, H.~Ooguri and Y.~Oz,
``Large $N$ field theories, string theory and gravity'',
Phys.\ Rept.\  {\bf 323} (2000) 183,
{\tt hep-th/9905111}.

\bibitem{HS} 
M.~Henningson and K.~Skenderis,
``The holographic Weyl anomaly,''
JHEP {\bf 9807}, 023 (1998),
{\tt hep-th/9806087}.

\bibitem{AOY} O.~Aharony, Y.~Oz and Z.~Yin,
``M-theory on $AdS_p \times  S^{11-p}$  and superconformal field theories,''
Phys.\ Lett.\ {\bf B430}, 87 (1998),
{\tt hep-th/9803051};\\
S.~Minwalla,
``Particles on AdS(4/7) and primary operators on M(2/5) brane worldvolumes,''
JHEP {\bf 9810}, 002 (1998), 
{\tt hep-th/9803053};\\
R.G.~Leigh and M.~Rozali,
``The large N limit of the (2,0) superconformal field theory,''
Phys.\ Lett.\ {\bf B431}, 311 (1998),
{\tt hep-th/9803068};\\
E.~Halyo,
``Supergravity on $AdS(4/7) \times  S(7/4)$ and M branes,''
JHEP {\bf 9804}, 011 (1998),
{\tt hep-th/9803077}.

\bibitem{AF} G.~Arutyunov and S.~Frolov,
``Three-point Green function of the stress-energy tensor in the 
AdS/CFT  correspondence'',
Phys.\ Rev.\  {\bf D60} (1999) 026004,
{\tt hep-th/9901121}.

\bibitem{CFN} R.~Corrado, B.~Florea and R.~McNees,
``Correlation functions of operators and Wilson surfaces in the  
$d=6$, $(0,2)$ theory in the large $N$ limit'',
Phys.\ Rev.\  {\bf D60} (1999) 085011,
{\tt hep-th/9902153}.

\bibitem{BZ} F.~Bastianelli and R.~Zucchini,
``Three point functions of chiral primary operators in 
$d=3$, ${\cal N} = 8$ and  $d=6$, ${\cal N}=(2,0)$ SCFT at large $N$'',
Phys.\ Lett.\  {\bf B467} (1999) 61,
{\tt hep-th/9907047};
``Three point functions for a class of chiral operators in maximally  
supersymmetric CFT at large $N$'',
Nucl.\ Phys.\  {\bf B574} (2000) 107
{\tt hep-th/9909179};
``3-point functions of universal scalars in maximal SCFTs at large N'',
JHEP {\bf 0005} (2000) 047,
{\tt hep-th/0003230}.

\bibitem{BFT1} F.~Bastianelli, S.~Frolov and A.~A.~Tseytlin,
``Three-point correlators of stress tensors in maximally-supersymmetric  
conformal theories in d = 3 and d = 6'', 
Nucl.\ Phys.\  {\bf B578} (2000) 139,
{\tt hep-th/9911135};

\bibitem{BFT2} F.~Bastianelli, S.~Frolov and A.~A.~Tseytlin,
``Conformal anomaly of (2,0) tensor multiplet in six dimensions and 
 AdS/CFT correspondence'', JHEP {\bf 0002} (2000) 013,
{\tt hep-th/0001041}.

\bibitem{dis}
D.~M.~Capper and M.~J.~Duff,
``Trace anomalies in dimensional regularization'',
Nuovo Cim.\  {\bf A23} (1974) 173.

\bibitem{venti}
M.~J.~Duff,
``Twenty years of the Weyl anomaly'',
Class.\ Quant.\ Grav.\  {\bf 11} (1994) 1387
{\tt hep-th/9308075}.

\bibitem{dewitt}
B.S. De Witt in  ``Relativity, Groups and Topology'',
 edited by B.S. De Witt and C. De Witt, (Gordon Breach, New York,  1964);
 ``Relativity, Groups and Topology II'', edited by
B.S. De Witt and R. Stora (North Holland, Amsterdam, 1984).

\bibitem{Bir}
N.~D.~Birrell and P.~C.~Davies,
``Quantum Fields In Curved Space'',
(University Press, Cambridge, UK, 1982).

\bibitem{FrTs}
E.~S.~Fradkin and A.~A.~Tseytlin,
``Quantum properties of higher dimensional and dimensionally reduced 
supersymmetric theories'',
Nucl.\ Phys.\  {\bf B227} (1983) 252;
``Conformal anomaly in Weyl theory and anomaly free superconformal theories,''
Phys.\ Lett.\  {\bf B134} (1984) 187.

\bibitem{FB}
F.~Bastianelli,
``The path integral for a particle in curved spaces and Weyl anomalies'',
Nucl.\ Phys.\  {\bf B376} (1992) 113
{\tt hep-th/9112035};
F.~Bastianelli and P.~van Nieuwenhuizen,
``Trace anomalies from quantum mechanics,''
Nucl.\ Phys.\  {\bf B389} (1993) 53
{\tt hep-th/9208059};
F.~Bastianelli, K.~Schalm and P.~van Nieuwenhuizen,
``Mode regularization, time slicing, Weyl ordering and phase space path  
integrals for quantum mechanical nonlinear sigma models'',
Phys.\ Rev.\  {\bf D58} (1998) 044002
{\tt hep-th/9801105}.

\bibitem{Bon}
L.~Bonora, P.~Cotta-Ramusino and C.~Reina,
``Conformal anomaly and cohomology'',
Phys.\ Lett.\  {\bf B126} (1983) 305.
 
\bibitem{Bon2}
L.~Bonora, P.~Pasti and M.~Bregola,
``Weyl cocycles'',
Class.\ Quant.\ Grav.\  {\bf 3} (1986) 635.

\bibitem{kar}
D.~R.~Karakhanian, R.~P.~Manvelian and R.~L.~Mkrtchian,
``Trace anomalies and cocycles of Weyl and diffeomorphism groups'',
Mod.\ Phys.\ Lett.\  {\bf A11} (1996) 409
{\tt hep-th/9411068}.

\bibitem{WesZum} 
J.~Wess and B.~Zumino,
``Consequences of anomalous Ward identities'',
Phys.\ Lett.\  {\bf B37} (1971) 95.

\bibitem{DesSch}
S.~Deser and A.~Schwimmer,
``Geometric classification of conformal anomalies in arbitrary dimensions'',
Phys.\ Lett.\  {\bf B309} (1993) 279
{\tt hep-th/9302047}.

\bibitem{Ichi}
S.~Ichinose and N.~Ikeda,
``Weyl anomaly in higher dimensions and Feynman rules in coordinate space'',
J.\ Math.\ Phys.\  {\bf 40}, 2259 (1999)
{\tt hep-th/9810256}.

\bibitem{Erdmenger:1997gy}
J.~Erdmenger,
``Conformally covariant differential operators: properties and applications'',
Class.\ Quant.\ Grav.\  {\bf 14}, 2061 (1997)
{\tt hep-th/9704108}.

\bibitem{Anselmi:2000uk}
D.~Anselmi,
``Quantum irreversibility in arbitrary dimension'',
Nucl.\ Phys.\  {\bf B567}, 331 (2000)
{\tt hep-th/9905005}.

\bibitem{imb} 
C.~Imbimbo, A.~Schwimmer, S.~Theisen and S.~Yankelowicz,
``Diffeomorphisms and holographic anomalies'', 
Class.\ Quant.\ Grav.\  {\bf 17} (2000) 1129,
{\tt hep-th/9910267}.

\bibitem{cap}
A.~Cappelli and G.~D'Appollonio,
``On the trace anomaly as a measure of degrees of freedom'',
{\tt hep-th/0005115}.

\bibitem{ans3}
D.~Anselmi,
``Towards the classification of conformal field theories 
in arbitrary  dimension'',
Phys.\ Lett.\  {\bf B476}, 182 (2000)
{\tt hep-th/9908014}.

\bibitem{Anselmi:1999xk}
D.~Anselmi,
``Anomalies, unitarity, and quantum irreversibility'',
Annals Phys.\  {\bf 276}, 361 (1999)
{\tt hep-th/9903059}.

\bibitem{PR} 
T.~Parker and S.~Rosenberg,
``Invariants of conformal Laplacians",
 J.\ Diff.\ Geom.\ {\bf 25,} 199 (1987).  

\end{thebibliography}
\end{document}